\documentclass[preprint]{aastex}

\slugcomment{Submitted in Icarus,}

\begin{document}


\title{A proof that tidal heating in a synchronous rotation is always larger 
that in an asymptotic nonsynchronous rotation state}


\author{B. Levrard \footnote{also at Universit\'e de Lyon 1, Centre de Recherche Astrophysique de Lyon, Ecole Normale Sup\'erieure de Lyon, CNRS, 46 all\'ee d'Italie, F-69364 Lyon Cedex 07}}
\affil{Astronomie et Syst\`emes Dynamiques, IMCCE-CNRS UMR 8028, 
\\ 77 Avenue Denfert-Rochereau, 75014, Paris
\\
\email{blevrard@imcce.fr}}

\vspace{10cm}

\noindent Manuscript pages: 9\\
Figures: 0\\
Tables: 0\\



\clearpage

\vskip5cm
\noindent
Running head: Tidal heating at arbitrary eccentricity and obliquity
\vskip5cm
\noindent
Corresponding author:\\
\vskip1cm
\noindent
Benjamin Levrard\\
IMCCE/Observatoire de Paris\\
77 Avenue Denfert-Rochereau\\
75014 Paris, France\\
blevrard@imcce.fr\\
Phone: (33)140512132\\
FAX:(33)140512055\\

\clearpage

\noindent
{\bf Abstract:}
\vskip2cm
In a recent paper, Wisdom (2007, Icarus, in press) derived concise expressions for 
the rate of tidal dissipation in a synchronously rotating body for arbitrary orbital
eccentricity and obliquity. He provided numerical evidence than the derived rate is
always larger than in an asymptotic nonsynchronous rotation state at any obliquity and
eccentricity. Here, I present a simple mathematical proof of this conclusion and 
show that this result still holds for any spin-orbit resonance.




\newpage

\section{Introduction}

Tidal heating is a source of energy which can have a strong influence on the
thermal and internal history of celestial bodies.  
For solid bodies like the terrestrial planets or rocky satellites,  
local heating is presumed to arise from the conversion of the mechanical strain energy 
associated with time-dependent tidal distortion. 
This may occur when a rotating satellite librates on an eccentric orbit or has a non-zero obliquity.

For a homogeneous and incompressible synchronously rotating satellite, the expression 
of tidal heating as a function of the eccentricity has been calculated in detail by Peale and Cassen (1978). 
The derivation is made by calculating the power dissipated by the tide-raising force
on each internal displaced constituent of the satellite. More recently, 
Wisdom (2004) generalized this expression to the second order in obliquity. 

For giant gaseous planets that are not expected to be trapped in spin-orbit
resonances but rather to reach an asymptotic nonsynchronous state, Levrard et al.(2007)
argued that previous expressions are inadequate in this situation and derived new
expressions of tidal dissipation for an asymptotic nonsynchronous rotation valid
at arbitrary eccentricity and obliquity. 
Unfortunately, they compared their results to the expressions given in Wisdom (2004) and
concluded that the rate of tidal heating in synchronous rotation is always lower than
in an asymptotic nonsynchronous state. Using new derivations and useful formulae for the rate of tidal heating 
in a synchronously rotating body valid for any eccentricity and obliquity, Wisdom (2007)
found the opposite result.

In this very short note, I provide a mathematical proof of this conclusion. Because solid (exo)planets and satellites are also expected eventually to despin to a more general state of spin-orbit resonance where the orbital period is some integer or half-integer times the rotation period (e.g. Goldreich and Peale 1966; Dobrovolskis 2007), I generalize
this result to all the other spin-orbit resonances.

\section{Comparison of tidal dissipation between a synchronously and asymptotically nonsynchronous
rotating body}

We consider the gravitational tides raised by an host planet on a
satellite (the demonstration also holds for a planet around its central star). 
We use the simplest model of tidal response, generally called ``viscous'' model 
as described in Mignard (1980).
He assumed a constant time lag for any frequency component of the tidal perturbation  
In other words, the tidally deformed surface of 
the satellite always assumes the equipotential surface it would have
formed a constant time lag $\Delta t$ ago, in the absence of dissipation.
In this case, the ratio $1/Q$ where $Q$ is the satellite effective tidal dissipation factor
is proportional to the frequency of the tides.

Here, I calculate the rate of tidal dissipation from the variation of the mechanical energy   
(rotational + orbital) of the satellite caused by tidallly-driven perturbations
in the satellite's rotational and orbital parameters (e.g. Hut 1981).
In that case, the total energy is
\begin{equation}
E=\frac{1}{2} C \omega^2-\frac{G M_s M_p}{2\,a},
\label{energy}
\end{equation}
where $\omega$ is the satellite's rotation rate, $C$ is its polar moment of inertia, $M_p$ the mass
of the host planet (primary body), $M_s$ the satellite mass, and $a$ is the orbit semimajor axis.

For a satellite locked into a synchronous resonance (1:1),
it is necessary to add to the right hand side of equation (\ref{energy}), the external gravitational potential 
of the deformed satellite caused by its permanent quadrupole moment. Averaged over an orbital period, it is classically
given by \footnote[1]{For simplicity, a zero obliquity is assumed but the same conclusion holds at any obliquity.}
\begin{equation}
V=-\frac{3}{4}\,H(1,e)\,(B-A)\,n^2\, \cos 2\gamma \,
\label{pot}
\end{equation}
where $H(1,e)$ is the Hansen's coefficient for the synchronous resonance, $e$ is the orbital
eccentricity, $n$ is the orbital mean motion, $A$ and $B$ are the satellite's 
equatorial moments of inertia and $\gamma$ is the resonant angle with 
\begin{equation}
d\gamma/dt=\omega-n \,.
\label{res}
\end{equation}
The gravitational restoring torque exerted by the planet on the
quadrupole moment maintains the spin in the resonance.

Combining equations (\ref{energy}),(\ref{pot}) and (\ref{res}), the rate of tidal dissipation within the satellite then is
\begin{equation}
\dot{E}_{tidal}=-\frac{dE}{dt}=-C\omega \times \frac{d\omega}{dt}-
\frac{3}{2}\,H(1,e)\,(B-A)\,n^2\,\frac{d\gamma}{dt}\,\sin 2\gamma -\,\frac{G M_s M_p}{2\,a^2} \times \frac{da}{dt}\,.
\label{energy2}
\end{equation}
Using the equation for the rotational motion of the satellite averaged over an orbital period (e.g. Murray \& Dermott 1999)  :
\begin{equation}
C \frac{d^{2}\gamma}{dt^{2}}= C \frac{d\omega}{dt}=-\frac{3}{2}\,H(1,e)\,(B-A)\,n^2\, \sin 2\gamma \,+\Gamma_{tidal}\,
\label{danby}
\end{equation}
where $\Gamma_{tidal}$ is the mean tidal torque acting to brake the spin of the satellite, equation (\ref{energy2}) becomes
\begin{equation}
\dot{E}_{tidal}=-\omega\, \Gamma_{tidal}+\frac{3}{2}\,H(1,e)\,(B-A)\,n^3\,\sin 2\gamma -\,\frac{G M_s M_p}{2\,a^2} \times \frac{da}{dt}\,.
\label{energy3}
\end{equation}

For most of satellites and planets of the solar system, the libration period is much longer than the orbital
period but much shorter than the typical despinning timescale so that $\left < \sin (2 \gamma) \right > =0$ over a libration period.
Averaging the equation (\ref{energy3}) over a libration period or over secular timescales longer than the
libration period then leads to
\begin{equation}
\left \langle \dot{E}_{tidal} \right \rangle=-\omega \left \langle  \Gamma_{tidal}\right \rangle -\,\frac{G M_s M_p}{2\,a^2} \times \left \langle  \frac{da}{dt} \right \rangle\,.
\label{energy4}
\end{equation}

For the ``viscous'' tidal model, the average tidal torque is given by 
(e.g. Hut 1981; N\'eron de Surgy \& Laskar 1997):
\begin{equation}
\left \langle \Gamma_{tidal}\right \rangle=-\frac{K}{n}
  \left[\left(1+x^2\right)\Omega(e)\frac{\omega}{n}-2x\,N(e)
  \right]\,
\label{rot_tidal}
\end{equation}
where $\varepsilon$ is the satellite's obliquity (the angle between the satellite equatorial and orbital planes),
 $x=\cos \varepsilon$ and
\begin{equation}
K = \frac{3}{2} \frac{k_2}{Q_n} \left( \frac{G
M_s^2}{R_s} \right) \left(\frac{M_p}{M_s} \right)^2 \left( \frac{R_s}{a}
\right)^6 n\,
\end{equation}
where $k_2$ is the potential Love number of degree 2, $R_s$ is the satellite radius and 
$Q_n=(n\,\Delta t)^{-1}$ is the annual tidal quality factor.
The eccentricity-dependent functions $N(e)$ and $\Omega(e)$ are then 
$$\Omega(e) = \frac{1+3e^2+\frac{3}{8}e^4}{(1-e^2)^{9/2}}
$$
and 
$$
N(e) = \frac{1+\frac{15}{2}e^2+\frac{45}{8}e^4+\frac{5}{16}e^6}{(1-e^2)^{6}}\,.
$$
In the same way, we have (e.g. Hut 1981; N\'eron de Surgy \& Laskar 1997):
\begin{equation}
\left \langle \frac{da}{dt} \right \rangle=4\,a^2
\left(\frac{K}{G\,M_s\,M_p}\right)\left[N(e)\,x\,
\frac{\omega}{n}-N_a(e)\right] \ ,
\label{evol_a}
\end{equation}
where 
$$ N_a(e)=\frac{1+\frac{31}{2}e^2+\frac{255}{8}e^4+\frac{185}{16}e^6+\frac{25}{64}e^8}
{(1-e^2)^{15/2}}\,.$$

Substituting equations (\ref{rot_tidal}) and (\ref{evol_a}) into the equation (\ref{energy4}) provides
the average rate of energy dissipation
\begin{equation}
\left \langle \dot{E}_{tidal} \right \rangle=2\,K\left[N_a(e)+\frac{1+x^2}{2}\Omega(e)-2xN(e)\right]\,
\label{synch_ene}
\end{equation}
using $\omega \simeq n$ and valid to any order in eccentricity and obliquity.

We checked that this expression fully agrees with the equation (30) of Wisdom (2007) calculated
for a homogeneous, incompressible and small and/or rigid enough body that the radial displacement
Love number $h_2$ is $5k_2/3$. Note that our derivation does not require such an 
hypothesis and all the uncertainties in the radial distribution of material and
its physical properties (e.g. density, compressibility, elasticity) are lumped into the $k_2$ parameter.
For a homogeneous and incompressible body, the Love number of degree 2 is given by the well-known formula $k_2=(3/2)/\left(1+19\,\mu/(2\,\rho g R_s)\right)$ where $\rho$ is the density and 
$\mu$ is the elastic shear modulus. For real rocky material, 
the effect of compressibility can not be actually neglected because
$\mu$ is close to $\lambda$ where $\lambda$ is classically the Lame's parameter which
is a measure of compressibility ($\lambda \longrightarrow \infty$ for an incompressible material). 
However, if deviations from incompressibility and homogeneous density requires only small corrections, 
a varying elastic shear modulus could have important consequences (e.g. Peale and Cassen 1978).
For a homogeneous and incompressible fluid planet, $k_2= 3/2$ but is
about one order of magnitude smaller for realistic profiles of density. Futhermore,
when the strong effect of compressibility is taken account, it is worthy to note that 
Love numbers could experience dramatic and unexpected variations (e.g. Hurford et al. 2002).

Let us now consider the same satellite (or planet) that is not locked into a synchronous resonance. 
This may occur because the body is essentially fluid or near-fluid and does not have a
permanent quadrupole moment, or because its eccentricity is too large to allow a capture into 
a synchronous resonance (e.g. Goldreich and Peale 1966). 

Because the satellite's orbital angular momentum is generally much larger than
its spin angular momentum, gravitational tides affect the spin's properties
(rotation, obliquity) much faster than the orbit's properties.
For the ``viscous'' model, tides ultimately reduce the obliquity to zero 
on the same time scale as the despinning (e.g. Hut 1981).
Here, we consider that other separate mechanisms may maintain
or excite the obliquity to a non-zero value (due for exemple to a capture in a Cassini state, to perturbations by
a companion, ...).
Setting $\left \langle \Gamma_{tidal} \right \rangle=0$ in the equation (\ref{rot_tidal}),  
the spin evolves to its asymptotic equilibrium rate of rotation
\begin{equation}
\omega_{eq}=\frac{N(e)}{\Omega(e)}\frac{2x}{1+x^2}\,n \,
\label{rot_eq}
\end{equation}
while the eccentricity and the obliquity are assumed to vary more slowly.
Note that a non-zero eccentricity favors a supersynchronous rotation (that is, a rotational
period shorter than the orbital period) while a non-zero obliquity favors a subsynchronous rotation. 
As a consequence, the asymptotic rate of rotation can be higher or lower than 
the synchronous rotation rate depending on the eccentricity and obliquity values.

Once the spin reaches its ``pseudo-equilibrium'', tidal energy is then dissipated within the satellite only at 
the expense of the orbital energy. From equation (\ref{energy4}), it comes 
$$\left \langle \dot{E}_{tidal} \right \rangle =-(G M_p M_s)/(2\,a^2) \times 
\left \langle da/dt \right \rangle.$$
The derived rate of tidal dissipation is then (see Levrard et al. 2007)
\begin{equation}
\left \langle \dot{E}_{tidal} \right \rangle=
2K\left[N_a(e)-\frac{N^2(e)}{\Omega(e)} 
\frac{2x^2}{1+x^2}\right].
\label{energy_ns}
\end{equation}

It is now easy to compare the rate of tidal dissipation in a synchronously rotating body
(Equation (\ref{synch_ene})) with that in an asymptotic nonsynchronous rotation rate 
(Equation (\ref{energy_ns})) for any eccentricity and obliquity. 
The former is larger than the latter if 
\begin{equation}
\frac{1+x^2}{2}\Omega(e)-2xN(e) \geq -\frac{N^2(e)}{\Omega(e)} 
\frac{2x^2}{1+x^2}\,.
\end{equation}
We found that this condition is always verified because the previous
equation is equivalent to 
\begin{equation}
[(1+x^2)\Omega(e)-2xN(e)]^2 \geq 0\,.
\end{equation}

As a consequence, the dissipation rate in synchronous rotation is always equal to
or larger than that in asymptotic rotation for any obliquity and eccentricity.
Note that the equality holds when the asymptotic rate of rotation is
synchronous, that is the satellite's eccentricity and obliquity verify
\begin{equation}
\frac{N(e)}{\Omega(e)}\frac{2x}{1+x^2}=1 \,.
\end{equation}
For a zero obliquity, it is possible only if the eccentricity is zero.
The function $N(e)/\Omega(e)$ is monotonic and always increases with the eccentricity 
so that there is only one eccentricity and obliquity
value that satisfies the previous condition.

\subsection{Generalization to other spin-orbit resonances}
Let us now assume that the satellite is locked into another spin-orbit
resonance. We have $\omega \sim p\, n$ where $p$ is an integer or
a half-integer. Using equations (\ref{energy4}), (\ref{rot_tidal}) and (\ref{evol_a}), the rate of tidal
dissipation for a resonant rotation is then
\begin{equation}
\left \langle \dot{E}_{tidal} \right \rangle=2\,K\left[N_a(e)+p^2\,\frac{1+x^2}{2}\Omega(e)-2\,p\,xN(e)\right].
\end{equation}
This rate of dissipation is larger than the asymptotic non-resonant rate of rotation if
\begin{equation}
p^2\,\frac{1+x^2}{2}\Omega(e)-2\,p\,xN(e) \geq -\frac{N^2(e)}{\Omega(e)} 
\frac{2x^2}{1+x^2}\,.
\end{equation}
Once again, this condition is always verified because this relationship is also equivalent to
\begin{equation}
[p(1+x^2)\Omega(e)-2xN(e)]^2 \geq 0\,.
\end{equation}
The equality holds when the asymptotic rate of rotation is equal to the rotation rate in the
spin-orbit resonance that is
\begin{equation}
\frac{\omega}{n}=\frac{N(e)}{\Omega(e)}\frac{2x}{1+x^2}=p \,. 
\end{equation}
Only one eccentricity and obliquity verify this equality.

\section{Conclusion}

We provide a mathematical demonstration that the rate of tidal dissipation in a 
synchronously rotating satellite is larger than that in a asymptotic nonsynchronous
rotation state for any obliquity and eccentricity. 

We also show that this property still holds for the 
other spin-orbit resonances. It would be interesting to
investigate whether these results are also valid for other tidal models.

\acknowledgments

I thank reviewers Jack Wisdom and Anthony Dobrovolskis for helpful reviews and 
stimulating discussions.

\clearpage

\clearpage







\clearpage

\end{document}